\documentclass[12pt, draftclsnofoot, onecolumn]{IEEEtran}

\usepackage{graphicx}
\usepackage{multirow}
\usepackage{subfigure}
\usepackage{amsmath}
\usepackage{amssymb}
\usepackage{mathptmx}
\usepackage{amsfonts}
\usepackage{times}
\usepackage{color}
\usepackage{fancyhdr}
\usepackage{gensymb}
\usepackage{lineno}
\usepackage{booktabs}
\usepackage{eurosym}
\usepackage[table,xcdraw]{xcolor}

\usepackage{hyperref}
\usepackage{color}

\usepackage{colortbl}
\usepackage{bm}

\usepackage{soul}

\begin{document}

\title{Bluetooth 5:  a concrete step forward towards the IoT}


\author{Mario~Collotta,~\IEEEmembership{Member,~IEEE,}~Giovanni~Pau,~\IEEEmembership{Member,~IEEE,}~Timothy Talty,~\IEEEmembership{Senior Member,~IEEE}~and~Ozan K. Tonguz,~\IEEEmembership{Senior Member,~IEEE}
\IEEEcompsocitemizethanks{
\IEEEcompsocthanksitem 
M. Collotta and G. Pau are with the Faculty of Engineering and Architecture\protect\\
Kore University of Enna, 94100 - Enna, Italy.\protect\\
E-mail: \{mario.collotta;giovanni.pau\}@unikore.it
 }
  \IEEEcompsocitemizethanks{
\IEEEcompsocthanksitem 
T.Talty is with General Motors, Detroit, Michigan, USA\protect\\
E-mail: timothy.talty@gm.com
 }
  \IEEEcompsocitemizethanks{
\IEEEcompsocthanksitem 
O.K. Tonguz is with the Departement of Electrical and Computer Engineering\protect\\
Carnegie Mellon University, Pittsburgh, PA 15213-3890, USA.\protect\\
E-mail: tonguz@ece.cmu.edu
 }
\thanks{}
}


\IEEEcompsoctitleabstractindextext{%
\begin{abstract}

Six years after the adoption of the standard 4.0, the Bluetooth Special Interest Group (SIG), a non-profit association that deals with the study and the development of technology standards including those of Bluetooth, has officially released the main features of Bluetooth 5.0. It is one of the significant developments in short-range wireless communication technology. As stated by the SIG, the new standard will forever change the way people approach the Internet of Things (IoT), turning it into something that takes place around them in an almost natural and transparent way. In this article, the future IoT scenarios and use cases that justify the push for Bluetooth 5 are introduced. A set of new technical features that are included in Bluetooth 5 are presented, and their advantages and drawbacks are described.

\end{abstract}

\begin{IEEEkeywords}
Bluetooth 5; Internet of Things; Bluetooth Low Energy.
\end{IEEEkeywords}}

\maketitle

\IEEEdisplaynotcompsoctitleabstractindextext

%
\IEEEpeerreviewmaketitle

\section{Introduction}
\label{sec:introduction}

Bluetooth is a technology that has been developed more than twenty years ago. It was 1994 when the first draft of data transmission that would become part of the digital modem ecosystem was presented. Nowadays, this technology is one of the pillars of the Internet of Things (IoT) \cite{7000963}, a worldview that provides continuous connections among technological accessories to achieve significant impact and improved performance \cite{7253713}. Bluetooth is used for data transmission through radio waves, which allows two or more devices to connect with each other. There is no limit to the type of file that can be transmitted, as such files could comprise data collected by sensors, photos, documents, music, and videos. However, the maximum range of a device equipped with Bluetooth does not usually exceed 100 meters, although there are three distinct classes, according to the signal coverage range. Class 1, the most powerful, can reach up to 100 meters; Class 2, the most common, operates only within 10 meters; Class 3, on the other hand, does not exceed 1 meter and is also the least used, especially recently.

In 2003 the Bluetooth technology was considered to be dead \cite{dead}, but, fortunately, that did not happen because Bluetooth has enjoyed tremendous popularity and prosperity in the last ten years in several application areas, such as audio communications and stereo streaming. The Bluetooth industry is thriving now and working on expanding the implementation of the technology to short-range wireless communication markets, other than audio and stereo communications, such as the IoT and Machine-to-Machine (M2M) communications \cite{7516570}. This goal can be accomplished thanks to the maturity of the technology and its strong presence in the marketplace.

However, for Bluetooth to be suitable for M2M and IoT applications \cite{6963809}, it needs to reduce the power consumption \cite{s151024818}, so that it can be used in battery-powered devices for a more extended period. In fact, IoT devices have to be connected, but power consumption usually is a real concern. For instance, regarding wearable devices, such as fitness wristbands, the time between recharging could make or break the commercial viability of a product. Even though Wi-Fi technology looks attractive for IoT, the system developers have quickly realized that the power consumption associated with Wi-Fi technology is too high, leading to a short active time before recharging.


The Bluetooth Special Interest Group (SIG) introduced the Bluetooth Low Energy (BLE), also called Bluetooth Smart \cite{7347955}, to lure application developers of emerging IoT. BLE was first specified in Bluetooth 4.0 and further improved in Bluetooth 4.1 and 4.2 \cite{bt4all}. Besides, BLE was shown to be an attractive technology for many applications including vehicular networks \cite{7010544}. Recently, the Bluetooth SIG has presented the specifications of Bluetooth 5 \cite{bt5release}, whose primary purpose is to offer significant enhancements compared to the preceding specification, regarding the range, speed, and broadcasting capacity. In the twisted battle for the control over the IoT communication standards, these new improvements might help BLE to prevail and to become the ultimate standard for IoT. This article presents an overview of the new improvements introduced by Bluetooth 5 and explains the potential advantages of each one. An analysis on how these enhancements will enable current and future applications is presented, showing how Bluetooth 5 will be a competitive technology in IoT applications.

\section{New Features and Concepts}
\label{sec:newfeatures}

Unlike previous iterations of the Bluetooth standard, released as “.0” and followed by updates (such as 4.1 and 4.2), the new standard is known merely as Bluetooth 5. The classic version of Bluetooth 5 is identical to the previous versions, while the significant innovations focus on the BLE version. According to the specification introduced by the SIG, hardware boards can support three types of Bluetooth connections \cite{bt5release}. They are BLE 4.x, Bluetooth 5 at 2 Mbps, and Bluetooth 5 Coded. BLE 4.x is the connection model used by the BLE spec, i.e., 4.0, 4.1 and 4.2. This connection type is known as the BLE at 1 Mbps because that is its expected rate at the lowest layer before protocol overheads are added. Bluetooth 5 at 2 Mbps is the new high-speed connection that has been presented. In this case, at the PHY layer, its speed is 2 Mbps. Bluetooth 5 Coded is a new particular connection type that comes with Bluetooth 5. Its goal is to provide long-distance connections, but with a lower bit rate. So, the primary objective is a broader range rather than speed.

It is clear that the new standard has been designed to create a communication network that ensures, over a short distance, a communication bandwidth that allows data exchange among the connected appliances and other smart devices of the IoT. So, this implies a network for automation, which further stimulates the development and the deployment of “smart and interconnected objects.” The Bluetooth SIG summarizes the innovations that the new standard introduces, as compared to the previous versions of Bluetooth standard (see Table \ref{tab:btcomparison}), and they are presented in this section. Moreover, Table \ref{tab:btcomparison} compares the various versions of Bluetooth with two other wireless protocols that are competing to dominate the vast IoT market; i.e., IEEE 802.15.4/ZigBee and IEEE 802.11ah/HaLow. Both protocols may be attractive for the IoT because each one has unique features that might be desirable for different applications. Nevertheless, the final specifications of HaLow have been presented in 2016, and still few devices are available on the market. For this reason, in this paper, a direct comparison with IEEE 802.15.4, through a real testbed, will be performed in Section \ref{sec:performance}.

\begin{table}[]
	\centering
	\tiny
	\caption{Technical comparison of Bluetooth versions and other wireless standards.}
	\label{tab:btcomparison}
	\begin{tabular}{|c|c|c|c|c|c|}
		\hline
		\textit{\textbf{Feature}} & \textbf{Bluetooth Classic} & \textbf{Bluetooth 4.x} & \textbf{Bluetooth 5} & \textbf{IEEE 802.15.4 - ZigBee}   & \textbf{IEEE 802.11ah - HaLow} \\ \hline
		Radio Frequency (MHz)     & 2400 to 2483.5             & 2400 to 2483.5         & 2400 to 2483.5       & 868.3, 902 to 928, 2400 to 2483.5 & 900                            \\ \hline
		Distance/Range (meters)   & Up to 100                  & Up to 100              & Up to 200            & Up to 150                         & Up to 1000                     \\ \hline
		Medium Access Technique   & Frequency Hopping          & Frequency Hopping      & Frequency Hopping    & CSMA/CA                           & Restricted Access Window       \\ \hline
		Nominal Data Rate (Mbps)  & 1-3                        & 1                      & 2                    & 0.02-0.25                         & 0.15-7.8                       \\ \hline
		Latency (ms)              & \textless 100              & \textless 6            & \textless 3          & \textless 4                       &  \textasciitilde1000                    \\ \hline
		Network Topology          & Piconet, Scatternet        & Star-bus, Mesh         & Star-bus, Mesh       & Mesh                              & Star-bus                       \\ \hline
		Multi-hop Solution        & Scatternet                 & Yes                    & Yes                  & Yes                               & Up to 2 hops                   \\ \hline
		Profile Concept           & Yes                        & Yes                    & Yes                  & Yes                               & No                             \\ \hline
		Nodes/Active Slaves       & 7                          & Unlimited              & Unlimited            & Unlimited                         & Unlimited                      \\ \hline
		Message Size (bytes)      & Up to 358                  & 31                     & 255                  & 100                               & 100                            \\ \hline
		Certification Body        & Bluetooth SIG              & Bluetooth SIG          & Bluetooth SIG        & ZigBee Alliance                   & IEEE                           \\ \hline
	\end{tabular}
\end{table}

\subsection{Larger Range}
\label{subsec:range}

The first significant change is related to a meaningful increase in the covered range. Bluetooth 4.x has a range between 50 and 100 meters, as the crow flies, outdoor and unobstructed, which is reduced to 10/20 meters in indoor environments. Bluetooth 5 aims to quadruple the range of BLE devices. In fact, in the worst case, this range should be 200 meters outdoors and about 40 meters indoors. These values can result in significant savings in the electronics world. Bluetooth 5 proposes a particular connection type which has been developed for long distance communications. It is important to note that this connection type is not suitable for Bluetooth speakers or syncing smartphones/wearable devices. In fact, this operating mode aims at the IoT where it is necessary to place low-cost modules all over a building or in an open space and gather data. The obtained data can be anything from light, humidity, temperature, traffic monitors or movement detectors, and so on; in fact, the opportunities are countless. However, these sensors need to send their data to a central hub/gateway and, necessarily, require a power supply. The power becomes a non-issue if the device is connected to the a.c. power supply. In this case, for instance, the device could use Wi-Fi to communicate. Nevertheless, the requirement for energy consumption and Wi-Fi coverage limits the potential of such devices.

In this context, Bluetooth 5 might have an edge since the devices that use Bluetooth do not necessarily require mains power. They could be installed with a simple battery. A possible solution to extend the range without increasing the power consumption is to decrease the data rate. An impressive feature of the PHY innovations in Bluetooth 5 is the way for enhancing the sensitivity. It is realized only in the LE Coded connection type. Typically, in BLE the packet headers and payloads are un-coded, i.e., 1 bit refers to 1 modulated symbol. This feature remains unchanged for both 1 Mbps and 2 Mbps connection modes. However, in the LE Coded connection method introduced in Bluetooth 5, two network data rates can be used: 500 kbps and 125 kbps. In these cases, the payloads have many symbols for each bit, S=2 for 500 kbps and S=8 for 125 kbps, where S is the symbol/bit rate. More symbols per bit translate to an increased tolerance with a weak Signal to Noise Ratio (SNR), and still provide a recoverable data stream. This mechanism occurs entirely in hardware in a transparent way to the developer. The coding process involves two stages. The first is the Forward Error Correction (FEC), and, afterward, a pattern mapper outlines a bit code to the input bits. The effect of these steps is a spreading of the data that provides a recovery using the FEC if bit errors occur and an improved ability to recover the received bit stream. This goal can be reached in conditions where the SNR is reduced to a level that data recovery would be impracticable without the LE Coded mode.

A four-fold range means a quadruple decrease in the number of access points and range extenders, which in the IoT realm represents an essential capability in keeping all the equipment and nodes in an IoT network connected. Moreover, the increased capacity can allow creating, in a smoother way, communication channels with beacons located in many different places; those open to the public, for instance in an airport, in a train station, in a shopping mall, or for the use in the home. In this way, wireless connectivity, increasingly used to provide location-based services, with fewer constraints than current technology, can be fully available. The extended range means that Bluetooth 5 might be able to replace Wi-Fi as a communication technology for many IoT applications.

\begin{figure}
	\centering
	\includegraphics[width=5in]{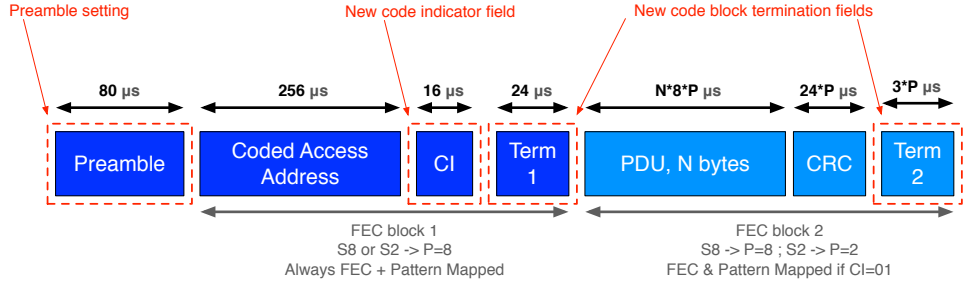}
	\caption{Packet structure with LE Coded mode.}
	\label{fig:packet}
\end{figure}

Finally, regarding the range, the question focuses on the choice of the 125 kbps or 500 kbps data rate. It is evident that this decision is undoubtedly related to the requirements of the specific application. In fact, with 500 kbps it is feasible to achieve about twice the range of standard BLE at 1 Mbps, whereas, with 125 kbps twice the range of 500 kbps can be reached. Nevertheless, considering the packet structure when LE Coded mode is used, shown in Figure \ref{fig:packet}, a notable part of the packet is the preamble address, and coding is at 125 kbps all the time, even when selecting 500 kbps which occur in the 1 bit Cl (Connection Interval) field of the packet. In fact, for considerably simple sensor/actuator operations, for instance, from 4 and up to 8 bytes, it is not feasible to save that much at 500 kbps over 125 kbps concerning collision avoidance or power consumption, while the broader range is sacrificed. In conclusion, for simple sensor/actuator operations, it is preferable to use 125 kbps to obtain the extra range, while for higher data transmissions of tens of bytes or more, the data rate of 500 kbps can be employed to achieve its advantages.

\subsection{Greater Speed}
\label{subsec:speed}

The improvement in the data transfer speed is a significant feature. While Bluetooth 4.x can reach a maximum speed of 1 Mbps, the maximum speed that Bluetooth 5 can support is 2 Mbps. This means that future wearable devices will synchronize with twice the speed of the current ones. For most applications, the speed of current Bluetooth standard for transferring data is sufficient. In many IoT applications, speed is not a significant issue. This is true for use cases that do not involve streaming. For instance, considering wearable devices, such as fitness wristbands, the amount of data to be transferred is pretty modest, and the currently supported BLE data rate is sufficient. Nevertheless, even for such wearable devices, higher transfer speed can allow faster software and firmware updates and improves users experience. According to ABI Research estimates \cite{abi}, over 371 million of Bluetooth beacons will be exchanged by 2020. Thanks to a higher data transmission capacity than the current Bluetooth 4.x, the new standard can involve more data transmitted by several smart devices, not just by the classic smartphones and tablets. As a consequence, also products used in automotive, home, business, and industrial applications will be able to exchange information with each other and with the cloud.

Bluetooth 5 consolidates the packet extension feature of Bluetooth 4.2, and by improving the on-air transfer speed, the obtainable network data throughput is doubled, about up to 1400 kbps. 
The data is transmitted faster, but the gap among the packets has not been reduced. Bluetooth 5 is about 1.7 times faster than BLE 4.2. Another significant advantage of having a 2 Mbps data rate is that energy savings become possible. 

\subsection{Beacons Everywhere}
\label{subsec:beacons}

Another notable enhancement that comes with Bluetooth 5 is the extended broadcast capacity compared to previous versions. This improvement will have a meaningful result in the world of beacons and will grow the range of use cases for them. Proximity devices and beacon systems are accessories and chipsets capable of automatically sending, to all neighboring devices, localized information, such as restaurant menus, promotions, data on road traffic and much more. 

Bluetooth 5 also improves the capability of sending special data packets, called "advertising packets." This communication strategy allows two Bluetooth devices to exchange packets and information even if they are not synchronized with each other. The advertising packets, for instance, enable the device to scan nearby areas and find out the name of other devices with active Bluetooth technology and capability. In the new standard, the packets are larger, and this makes it possible to send more information also to asynchronous devices. This feature is fundamental to the development of an IoT network.

So, what distinguishes Bluetooth 5 from its predecessors is the manner in which it handles the beacons, i.e., messages/packets that can be transmitted continuously from a “presenting” device and received by nearby users on their smartphones, tablets or wearable devices. A Bluetooth 5 beacon does not require the pairing that comes with speakers, headsets, and other Bluetooth accessories. Considering that the beacon can broadcast information when a Bluetooth-enabled device arrives into range, site-specific content, location-enabled marketing, process updates, and other applications become possible. 

With Bluetooth 4.x, beacons can send messages of 31 bytes. This message size is small, considering that these bytes include both the message and any additional protocols to indicate data types in the package. Both Apple, with iBeacon, and Google, with Eddystone, have tried to get around this limitation by using the UUID (Universally Unique Identifiers), values of 128-bit capable of allowing the receiving devices to find a different beacon. However, when it is necessary to introduce more substantial information, such as URLs or telemetry data, it is clear that 31 bytes are not enough. Bluetooth 5 resolves this issue by increasing the size of the messages from 31 bytes to 255 bytes. 

Mesh networking support is another feature that is supported by Bluetooth 5, as previously stated by the SIG \cite{mesh}. However, although the Bluetooth 5 specifications have been released \cite{bt5release}, the mesh support is not included within them. In fact, the mesh networking for Bluetooth, known as Bluetooth Mesh, has been released only recently. 

\subsection{Longer Battery Life}
\label{subsec:battery}

\begin{figure}
\centering
\includegraphics[width=5in]{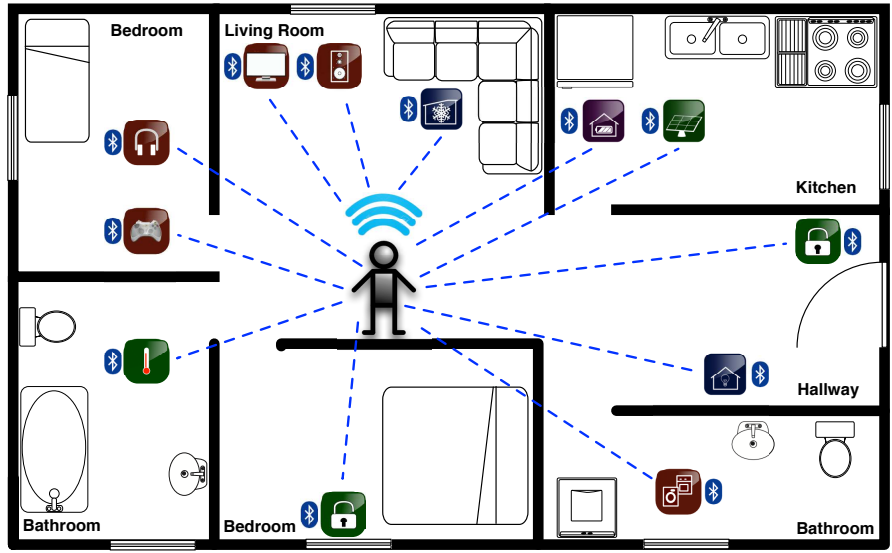}
\caption{Bluetooth-based home automation system.}
\label{fig:smarthome}
\end{figure}

The extended range and speed could imply a higher energy consumption. However, thanks to some ingenious design, such as the manner in which the signal is modulated and the advancement in the use of the frequency spectrum, Bluetooth 5 uses fewer energy resources. In fact, in the best case, it can consume about two times less power than the previous version of Bluetooth. If the speed is doubled in a wired device, the direct consequence is doubling of energy consumption. Nevertheless, it should be noted that Bluetooth operates at 2.4 GHz and this radio frequency dictates the power consumption, not the data rate. Bluetooth 5 enables exchanging twice the amount of data, and the result is that the device consumes half the power to transfer the same data. 
It is relevant to note that Bluetooth needed to be as good or better than comparative IEEE 802.15.4 solutions (i.e., ZigBee) to become a real competing technology within living environments. For this reason, new BLE output power classes have been introduced in Bluetooth 5:
	\begin{itemize}
	\item LE Class 1: Max +20dBm; Min $>$+10dBm;
	\item LE Class 1.5: Max +10dBm; Min -20dBm;
	\item LE Class 2: Max +4dBm; Min -20dBm;
	\item LE Class 3: Max 0dBm; Min -20dBm.
\end{itemize}

When selecting a wireless standard for battery-powered IoT applications, providing the same amount of data at half the power is already a very significant advantage. Besides, there are also reduced infrastructure costs, since no access points or routers are required. As a consequence, Bluetooth tends to be easier on battery life than Wi-Fi. The overall result should be the ability to make small, robust IoT devices for industrial and consumer applications.

\section{Scenarios, Use Cases and Performance}
\label{sec:performance}

The improvements in the covered range, higher data rate, and the increase in the size of the advertising packet, are designed and manufactured precisely to encourage the dissemination and the adoption of the new standard in the IoT world.

In a smart home scenario (Figure \ref{fig:smarthome}), broader coverage, for instance, not only allows to use Bluetooth audio speakers without having the problem of not being able to move with the smartphone in the next room but also to improve and to make the communication faster among smartphones and smartwatches. Using a smartphone, tablet, or laptop in hand, homeowners can manage the lights, temperature, home appliances, door and window locks and security systems in their home, even outside the home walls. This purpose may be possible thanks to Bluetooth sensors, for temperature, lights, doors, windows and motion detection, deployed in the home. In the end, homeowners can monitor and manage everything, from their lighting and home security system to window and door locks, with user-friendly applications. Since most homeowners already have at least one Bluetooth compatible smartphone, smartwatch or tablet, they can do this with devices they are already familiar with and know how to use.


Bluetooth 5 can allow IoT devices, like smartwatches, to move away from the present paired app-device model and to operate independently. As a consequence, the handshake procedure (pairing), to link to a data source and to authorize to get the data, is no longer required. In this case, some devices might benefit from increased autonomy, as battery power and screen size of a watch or wearable are both limiting factors, often requiring the help of a paired device, which in most circumstances is a smartphone.

\begin{figure}
	\centering
	\includegraphics[width=5in]{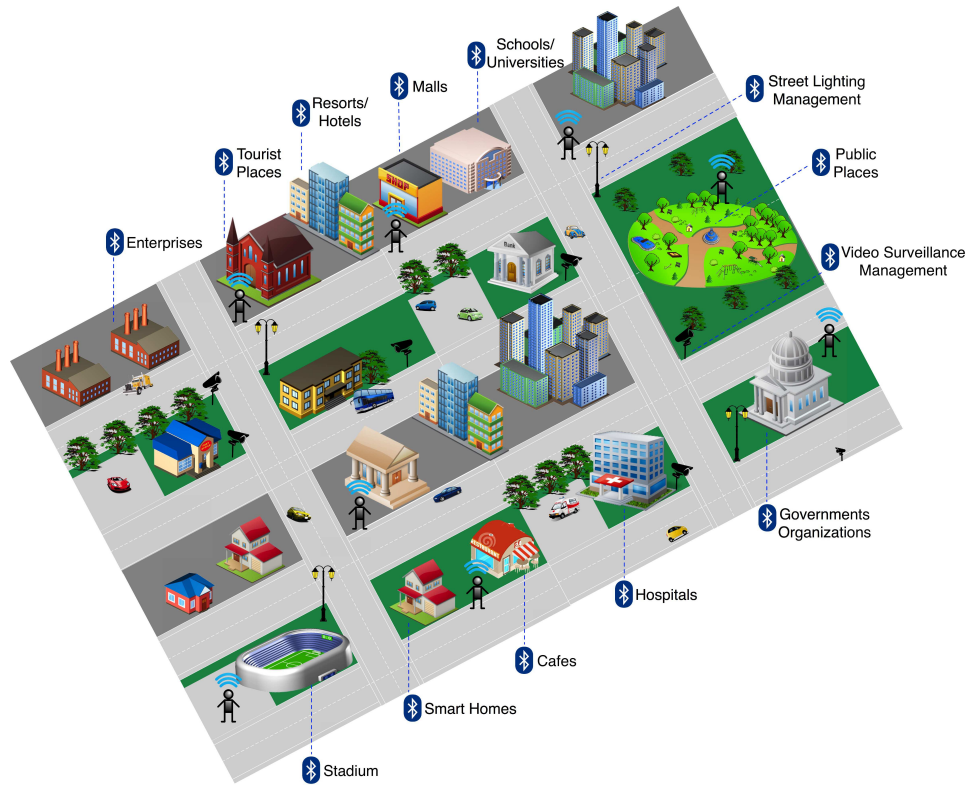}
	\caption{Bluetooth 5: Enabling Connected Smart City.}
	\label{fig:smartcity}
\end{figure}

In smart cities using IoT technologies (Figure \ref{fig:smartcity}), it is essential to employ an infrastructure that is intended to create hyper-location and automatic services. This context is precisely where the use of beacons come into play. Cities could deploy a mesh of beacons, on public or private networks, to collect and send data back to a centralized hub. The most significant advantage of using beacons is that, compared to other smart city technologies, they could be a lower cost solution. The use of Bluetooth 5 as a control mechanism also makes it possible for cities to control different types of equipment further improving efficiency remotely. In fact, Bluetooth-based solutions, for connected smart cities, can serve as the establishment for delivering next generation innovative services, for empowering the government, businesses, and citizens, thus enabling smarter cities. This transformation to smart cities can be achieved through an ecosystem that brings together solutions from different providers.

For instance, the managers of supermarkets, large retail chains, and the museum directors can also be interested in these technological developments. In the first case, beacons facilitate the expenditure of every day, guiding users within the various sectors and sending discount coupons and notifications about the latest offers based on their profile and their purchasing habits. All of this can be accomplished without the customer taking any actions. In museums,  for instance, Bluetooth can replace the audio guides. Visitors can just turn on the Bluetooth on the smartphone to receive information about the artworks. All this can be obtained, as mentioned previously, with a very low energy consumption.

Bluetooth 5 can be used as a street lighting control solution that offers lamp-level management capabilities of every street light in the city. This in-depth grid management can give an accurate real-time feedback on any change occurring along the grid, to reduce both energy loss and energy surges and offering advanced maintenance optimization tools. 


Another feature in Bluetooth 5 is the capability to broadcast richer data, including not only location information, but also URLs and multimedia files. For instance, multiple sensors in a store could provide internal GPS-style navigation to a specific item. The next time the consumer goes shopping, for example, there is no need to ask someone for directions. Bluetooth 5 could also be beneficial for self-driving cars, which have to interact with many sensors and external sources like traffic lights. As a direct result of beacons everywhere, with the arrival of Bluetooth 5, the IoT will benefit tremendously, despite the fact that several companies currently focus on the production of accessories and appliances based on Bluetooth 4.x connection. If Bluetooth 5 fulfills the promised expectations, it may be possible that even more vendors will decide to jump into the IoT market, mainly thanks to the much lower cost of the chipset.

Unfortunately, Bluetooth 5 does not allow to update old Bluetooth devices. In fact, the new version of Bluetooth requires a new type of chips that need to be installed on newer devices. The older versions of Bluetooth can work well on Bluetooth 5, but they do not have the same features., i.e., they will still work at their original speed and distance. 

\subsection{Bluetooth 5 Performance}
\label{subsec:btperf}

Several experimental measurements have been carried out to evaluate the performance of Bluetooth 5 compared to those of Bluetooth 4.2 and IEEE 802.15.4. In the testbed scenarios, Bluetooth 5 nRF52840 development boards \cite{nordic} from Nordic Semiconductor have been used. These boards implement a Bluetooth 5 protocol stack and a 32-bit ARM Cortex-M4F microcontroller clocked at 64 MHz. Moreover, they support three types of Bluetooth connections, i.e., BLE 4.x, Bluetooth 5 at 2 Mbps and Bluetooth 5 Coded (500 Kbps and 125 Kbps). On the contrary, regarding IEEE 802.15.4, prototyping boards equipped with Microchip PIC24FJ256GB108 microcontroller \cite{PIC24F} and MRF24J40MB radio frequency transceiver \cite{MRF24J40MB} both from Microchip Technology have been adopted. In both cases, the boards have been equipped with a 3 V coin cell battery. Utilizing this battery, its maximum level, when it is fully charged, is 250 mA, while the corresponding digital value, obtained through a 1O bit AD converter, is 1024.

A home network scenario has been investigated, and the performance has been evaluated in terms of battery consumption and throughput, both in indoor and outdoor contexts. Several nodes have been deployed in a fixed position, while a single device has been used as a mobile. The distance between this mobile board and the fixed ones has been changed between 20 and 5 meters in the indoor scenario, and between 35 and 120 meters in the outdoor scenario. The duration of the measurements has been 5 hours in each instance.

\begin{table}[]
	\centering
	\caption{Indoor/Outdoor scenario: battery consumption comparison.}
	\label{tab:battery}
	\begin{tabular}{|c|c|c|c|}
		\hline
		\multicolumn{2}{|c|}{\textbf{Wireless Protocol}}  & \textbf{\begin{tabular}[c]{@{}c@{}}Remaining battery\\ (digital value)\end{tabular}} & \textbf{\begin{tabular}[c]{@{}c@{}}Consumed battery\\ (percentage)\end{tabular}} \\ \hline
		Bluetooth 5   & \multirow{3}{*}{\textit{Indoor}}  & 707                                                                                  & 31                                                                               \\ \cline{1-1} \cline{3-4} 
		Bluetooth 4.2 &                                   & 645                                                                                  & 37                                                                               \\ \cline{1-1} \cline{3-4} 
		IEEE 802.15.4 &                                   & 0                                                                                    & 100                                                                              \\ \hline
		Bluetooth 5   & \multirow{3}{*}{\textit{Outdoor}} & 604                                                                                  & 41                                                                               \\ \cline{1-1} \cline{3-4} 
		Bluetooth 4.2 &                                   & 725                                                                                  & 29                                                                               \\ \cline{1-1} \cline{3-4} 
		IEEE 802.15.4 &                                   & 0                                                                                    & 100                                                                              \\ \hline
	\end{tabular}
\end{table}

Table \ref{tab:battery} shows that both Bluetooth 5 and Bluetooth 4.2 have obtained a significantly lower power consumption than IEEE 802.15.4. In fact, using the latter, the battery power has been exhausted even before the 5-hour time duration required for the experimental measurement. In the indoor case, Bluetooth 5 has achieved the best result. The same thing has not occurred in the outdoor context. In fact, the Bluetooth 5 Coded connection mode of the nRF52840 development board has a slightly higher power consumption than the BLE 4.x mode \cite{nordic}. Nevertheless, it is necessary to point out that, in this outdoor case, the range between the mobile device and the fixed one has reached up to 120 meters. This transmission range was not possible with Bluetooth 4.2 as the maximum range has been about 60 meters. A direct consequence of the transmission range is the measured throughput whose values are shown in Table \ref{tab:th}. These results prove that, in general, Bluetooth is faster than IEEE 802.15.4 and that Bluetooth 5 achieves better throughput performance than Bluetooth 4 in every case. Furthermore, for closer distances, the speed of Bluetooth 5 is significantly faster than BLE 4.2. Also, it is clear that the throughput decreases as the distance increases. In the indoor scenario, the Long Range mode of Bluetooth 5 has not been used and, as a result, as the range increases the advantages of Bluetooth 5 diminish compared to BLE 4.2, regarding the throughput. However, in the outdoor scenario with Line of Sight (LOS) communication, the maximum transmission range of Bluetooth 4.2 was about 60 meters. Data transmission by Bluetooth 5 not only has reached a range of about 120 meters but has achieved a better throughput than the other two wireless protocols even in this case.

In conclusion, our experimental measurements demonstrate the excellent improvements in Bluetooth 5. However, there are also dependencies on the underlying hardware capabilities of the radios used. In fact, four distinct data rates are available. Obviously, these are not achievable only through a firmware update because each device must support these features at the physical layer. The four improvements offered by Bluetooth 5 specs, i.e., doubling the speed, reducing the power consumption by 50\%, increasing the range of communication by a factor of four, and the boost in broadcasting, in conjunction with the potential ubiquity of the standard in consumer devices, should allow wider adoption of the standard, including the emerging IoT. As a result, it could be possible that Bluetooth 5 eventually becomes one of the fundamental wireless standards implemented in many IoT applications.

\begin{table}[]
	\centering
	\caption{Indoor/Outdoor scenario: throughput comparison.}
	\label{tab:th}
	\begin{tabular}{|c|c|c|c|c|}
		\hline
		\multicolumn{5}{|c|}{\textit{Indoor}}                                                                                                                                                                                                                                                                                         \\ \hline
		\textbf{\begin{tabular}[c]{@{}c@{}}Distance\\ (meters)\end{tabular}} & \textbf{Walls} & \textbf{\begin{tabular}[c]{@{}c@{}}Bluetooth 5\\ throughput\end{tabular}} & \textbf{\begin{tabular}[c]{@{}c@{}}Bluetooth 4.2\\ throughput\end{tabular}} & \textbf{\begin{tabular}[c]{@{}c@{}}IEEE 802.15.4\\ throughput\end{tabular}} \\ \hline
		20                                                                   & 4              & 485                                                                       & 380                                                                         & 95                                                                          \\ \hline
		15                                                                   & 2              & 584                                                                       & 533                                                                         & 130                                                                         \\ \hline
		11                                                                   & 2              & 910                                                                       & 635                                                                         & 140                                                                         \\ \hline
		5                                                                    & 1              & 1210                                                                      & 675                                                                         & 180                                                                         \\ \hline
		\multicolumn{5}{|c|}{\textit{Outdoor}}                                                                                                                                                                                                                                                                                        \\ \hline
		120                                                                  & LOS            & 105                                                                       & 0                                                                           & 85                                                                          \\ \hline
		80                                                                   & LOS            & 278                                                                       & 0                                                                           & 107                                                                         \\ \hline
		60                                                                   & LOS            & 577                                                                       & 488                                                                         & 125                                                                         \\ \hline
	\end{tabular}
\end{table}

\section{Discussion}
\label{sec:discussion}

IoT is a paradigm shift in coping with human mobility and connectivity. It is anticipated that IoT will provide the necessary ambient intelligence in every environment for handling the needs of people for a substantially improved way of life. Different industry forecasts are projecting 50 billion devices connected to the Internet by the year 2020. This estimate is an order of magnitude larger than the number of people who are connected to the Internet.

This vision, however, has a myriad of research and technology challenges that have to be met with viable engineering solutions. Below, we highlight some of the most important technical challenges of the future IoT:

\begin{itemize}
	\item \textbf{Scalability}: The sheer number of devices and users and the interactions required between them make scalability a major challenge.
	\item \textbf{Interoperability}: The heterogeneity of enabling devices and platforms make interoperability another key challenge in the IoT.
	\item \textbf{Efficiency in communications}: To have a viable IoT, low power sensors, wireless transceivers, communications, and networking for M2M will be crucial for efficient communications.
	\item \textbf{Security and Privacy}: Huge volumes of data emerging from the physical world, M2M, and new communications in the future IoT implies another major challenge which involves mining the data, providing secure access, and preserving privacy of the data.
	\item \textbf{Timeliness and Freshness of Data}: Ensuring the timeliness and freshness of data is another important challenge that has to be met.
	\item \textbf{Mobility, access, and service continuity}: Ubiquity for the IoT brings with it the challenge of addressing mobility, ad-hoc, access and service continuity.
	\item \textbf{Practical naming, resolution, and discovery}: The envisioned global access and discovery in the IoT requires practical naming, resolution, and discovery solutions as well.
\end{itemize}

Indeed, this is not a comprehensive list of all the challenges associated with the IoT, but it captures some of the most obvious and important ones. To put things into perspective, the main contribution of this paper is to show that Bluetooth 5 can address the third challenge mentioned above; namely, the efficiency of communications which is fundamental for a viable IoT. It would be interesting to explore how Bluetooth 5 fares regarding the other challenges mentioned above (such as scalability, interoperability, data mining, secure access, ubiquity). Such an investigation would lead to a more rounded global viewpoint, and it is the current direction of our research.

\section{Conclusions}
\label{sec:conclusions}

Bluetooth 5 aims to offer significant performance improvements compared to the previous versions of Bluetooth, regarding speed, range, and broadcasting capacity. In the fierce competition for dominating the IoT communication standard, these new advantages might help BLE to be one of the best choices for IoT. Bluetooth 5 quickly attracted the attention of investors, especially start-ups and venture capital firms, who look with interest to the burgeoning market of the IoT. Currently, it is hard to predict what will be the adopted wireless standard in the IoT. In fact, the dynamic and evolving world of smart and connected things is still in its infancy. However, considering the significant improvements in speed, power consumption, range, and capacity, it seems like Bluetooth 5 is a strong candidate.

\bibliographystyle{IEEEtran}
\bibliography{IEEEfull}

%




\begin{IEEEbiographynophoto}{Mario Collotta}
is a professor with tenure in the Faculty of Engineering and Architecture at the Kore University of Enna, Italy. He is the director of the Computer Engineering and Networks Laboratory. His research interests concern the realization of strategies and innovative algorithms to ensure a flexible management of resources in real-time systems and networks. He is a member of the IEEE and has published over 60 refereed papers in international journals and conferences.
\end{IEEEbiographynophoto}


\begin{IEEEbiographynophoto}{Giovanni Pau}
	is a professor at Faculty of Engineering and Architecture, Kore University of Enna, Italy. He has published more than 40 papers in journals and conferences and authored one book chapter. He serves as Associate Editor of several journals and serves/served as a leading Guest Editor in many special issues. His research interests include wireless sensor networks, soft computing techniques, internet of things, home automation, green communication, and real-time systems.
\end{IEEEbiographynophoto}


\begin{IEEEbiographynophoto}{Timothy Talty}
	joined Ford Motor Company in 1993, and worked on wireless channel modeling and concealed antenna systems development. He joined the EECS Department of the United States Military Academy, West Point, New York, as an assistant professor in 1997, where he conducted research on embedded antenna systems and high-speed Sigma-Delta converters. In 2001, he joined General Motors Corporation, Warren, Michigan, where he is currently a technical fellow working in the areas of wireless sensors and networks.
\end{IEEEbiographynophoto}


\begin{IEEEbiographynophoto}{Ozan K. Tonguz}
	is a tenured full professor in the Electrical and Computer Engineering Department of Carnegie Mellon University (CMU). He is the founder and CEO of the CMU startup known as Virtual Traffic Lights, LLC, which specializes in providing solutions to acute transportation problems using vehicle-to-vehicle (V2V) and vehicle-to-infrastructure (V2I) communications paradigms. His current research interests include vehicular networks, wireless and sensor networks, self-organizing networks, artificial intelligence, machine learning, smart grid, bioinformatics, and security. He currently serves or has served as a consultant or expert for several companies, major law firms, and government agencies in the United States, Europe, and Asia.
\end{IEEEbiographynophoto}





\end{document}